\documentclass[journal]{IEEEtran}
\usepackage{amsmath}
\usepackage{amssymb}
\usepackage{subcaption}
\usepackage{graphicx}
\usepackage{standalone}
\usepackage{tikz}
\usepackage{cite}
\usepackage{orcidlink}

\ifCLASSINFOpdf
\else
\fi

\begin{document}
\title{Analytical Formula for Calculations of Armour Losses in Three-Core Power Cables}
\author{Marius Hatlo
        and~Martin Hovde\orcidlink{0000-0002-3291-4996}

\thanks{Both authors have contributed equally to the work presented in this paper. \textit{(Corresponding author: M. Hovde)}}

\thanks{Marius Hatlo is with Unitech Power Systems AS, NO-1630 Fredrikstad, Norway (e-mail: marius.hatlo@unitech.no).}

\thanks{Martin Hovde is with Nexans Norway AS, NO-1788 Halden, Norway (e-mail: martin.hovde@nexans.com).}
}

\markboth{}%
{Shell \MakeLowercase{\textit{et al.}}: Bare Demo of IEEEtran.cls for IEEE Journals}
\maketitle

\begin{abstract}
Over the past decade, significant progress has been made in the field of loss and rating calculations for armoured three-core cables. This development was prompted by an industry realization that the applicable international standards often overestimate losses, leading to unnecessarily bulky and more expensive cables.
Starting with first-principles, this paper presents an accurate analytical formula for armour losses in three-core cables. The formula has undergone rigorous validation against 3D Finite Element Analysis (FEA) and demonstrate excellent accuracy. In the specific cases examined, the largest deviation from FEA results in terms of armour loss is approximately 2.4 percent for fully armoured cables. 
Although this study specifically focuses on armour losses, it establishes the groundwork for precise loss calculations in armoured three-core cables, including the conductor and screen losses. And the work presented here formed the basis for the complete loss calculations presented in the CIGRE Technical Brochure 908.
\end{abstract}


\IEEEpeerreviewmaketitle

\section{Introduction}
\IEEEPARstart{I}{n the last} decade, a significant amount of work has been published regarding losses in armoured three-core cables \cite{Bremnes2010,Hatlo2014,Hatlo2015,Goddard2015,Viafora2016,Sturm2015,Giussani2020,Giussani2021}. This surge in interest is partially driven by the expanding Offshore Windfarm (OWF) market, which demands large three-core cables.
Traditionally, IEC 60287 has served as the standard for these calculations, relying on formulas developed by Carter, Arnold, and other pioneers \cite{carter_1927, Arnold1941} . However, recent advancements aim to refine these methods and enhance their accuracy.

The formula for armour losses, as derived by Arnold in \cite{Arnold1941}, assumes that the armour wires are in perfect electrical contact, effectively acting as a shared screen for the three phases. To accommodate the magnetic characteristics of the armour, specific empirical factors were introduced to align with the measured data \cite{Arnold1941}. However, recent industry insights reveal that the underlying physical assumptions behind the IEC 60287 formulas do not hold for modern three-core cables.

Finite Element Analysis (FEA) results and measurements published in 2010 suggested that armour losses in three-core cables were significantly lower than predicted by the above-mentioned IEC 60287 standard \cite{Bremnes2010}. Discrepancy between measurements and IEC 60287 was mainly attributed to twisting of armour wires around the three conductors combined with poor electrical contact between individual armour wires, effectively cancelling induced circulating currents \cite{Bremnes2010}, and the associated Joule losses. Based on this, a 2.5D Finite Element Method was developed that was in good agreement with measurements performed on three-core cables carrying low currents \cite{Bremnes2010}.

Measurements published one year later on larger three-core cables carrying relatively large currents showed an increasing cable resistance with increasing current \cite{Karlstrand2011}. Similar trends were observed elsewhere in both measurements and FEA \cite{Hatlo2014, Sturm2015}. From this it was suggested that the observed increase in resistance with increasing current is due to a increase in permeability and hysteresis of the armour wire material with increasing magnetic field strength, a behavior that is typical in the Rayleigh region of magnetic materials\cite{Rayleigh1887,Hatlo2014}.

The realisation that circulating currents in the armour wires were eliminated by twisting was a major step forward \cite{Bremnes2010}. However, to fully understand the results of the measurements, the armour wire twisting and non-linear magnetic properties had to be combined with a magnetic field component oriented along the armour wires, which is a longitudinal effect not captured by the cross-sectional 2D and 2.5D FEA models \cite{Hatlo2014, Goddard2015,Hatlo2015,Maioli2015}. To add to the complexity, measurements of the magnetic properties of the armour wires showed a significant difference between the different armour steel grades \cite{Hatlo2014,Hatlo2015}. Further adding to the complexity, and in contradiction with already published results, new measurements on a large three core cable suggested that for certain cable designs the armour losses were in fact quite close to the IEC 60287 predictions \cite{Frelin2015}, and that the armour losses were nearly the same for a fully armoured cable compared to a half armoured cable, contrary to the common belief in the industry at the time. The apparent discrepancies and contradictions in the published measurements emphasised the need for a better understanding of the underlying physics. 

Numerical and analytical methods have been developed to approximate the armour losses, and the influence of the magnetic armour on the screen losses \cite{Goddard2015,Hatlo2015}. Research on the topic is still ongoing, and in recent years several papers have been published, on measurements\cite{Stolan2018}, 3D FEA simulations \cite{Lopez2018}, and numerical methods \cite{Giussani2020}, and a better understanding of the phenomenon has indeed been established in the industry. 

While the use of 3D FEA is getting increasingly more popular and efficient \cite{Lopez2018}, relevant FEA software are costly and in the general case only for the experienced user. It is therefore a need for simple analytical formulas that accurately describe cable losses. Whilst this paper does not include a model for calculating total cable losses, it presents a derivation of formulas for armour losses that can be implemented in e.g. the IEC 60287 framework for total loss calculations and cable rating, with the structure of the paper being as follows.

Firstly, we briefly introduce the model used for the magnetic auxiliary field generated by an un-armoured, un-screened three-core cable. This field will further serve as an exciting field in which the cable armour will be placed.

Secondly, as the exact problem with three twisted cores enclosed by $N$ twisted armour wires is too complex to tackle directly by analytical methods, we start off by transforming the $N$ wire armour to an equivalent tube representation. The idea of a tube representation is not novel as it is taken in e.g. IEC 60287 as well as in references \cite{Goddard2015} and \cite{Viafora2016}, but the specific approach taken herein is.

Thirdly, we solve Maxwell's equations for the twisted three cores surrounded by the equivalent tube armour. Then, from the solutions for the magnetic field $\mathbf{B}$ and the auxiliary field $\mathbf{H}$, the formula for the armour losses is derived.

Lastly, we validate the derived formulas with 3D FEA simulations, and investigate on approximations made in the derivations of the proposed formulae.
\section{Magnetic field generated by three twisted current-carrying filamentary conductors}
\begin{figure}[ht]
    \centering
    \begin{tikzpicture}[scale=0.25, every node/.style={scale=0.4}]
\node (O) at (0, 0) {};
\node (C) at (60.33mm, 0) {};
\def \R {231.2mm/2};
\node (X) at (55mm, 0) [scale=2, xshift=-1mm] {$x$};
\node (Y) at (0, 55mm) [scale=2, yshift=-1mm] {$y$};
\node (A) at (201:\R+3mm) {};
\foreach \x in {0, 3, ..., 360}{
    \filldraw[fill=white, draw=black] (\x:\R) circle [radius=5mm/2];
}
\foreach \x in {0, 120, 240}{
    \filldraw[fill=black, draw=black] (\x:60.33mm) circle [radius=1.5mm];
}
\draw[thick, ->] (0,0) -- node [scale=2, midway, left, yshift=-2mm, xshift=1mm] {$a_p$} (120:59.5mm);
\draw[thick, ->] (0,0) -- node [scale=2, midway, above, xshift=-1mm] {$R$} (A);
\draw[->] (0,0) -- (X);
\draw[->] (0,0) -- (Y);
\draw[-, shorten <=-2pt] (O) -- node [scale=2, midway, below, yshift=4mm, xshift=0mm] {$\rho$} (45:40mm);
\filldraw[] (45:40mm) circle [radius=0.1cm] node[above, scale=2, xshift=1mm]{$(\rho, \varphi)$};
\filldraw[] (O) circle [radius=0.1cm];
\draw (0:20mm) arc (0:45:20mm) node[midway, above, xshift=3mm, yshift=-4mm, scale=2]{$\varphi$};
\end{tikzpicture}
    \caption{Three symmetrically placed filamentary conductors depicted as dots carrying balanced three-phase r.m.s. currents $I_c$, enclosed by a ring of $N$ armour wires. The ring has a mean radius $R$, whilst the distance from the center of the cable to each filamentary conductor is $a_p$.}
    \label{fig:cable_fig_1}
\end{figure}
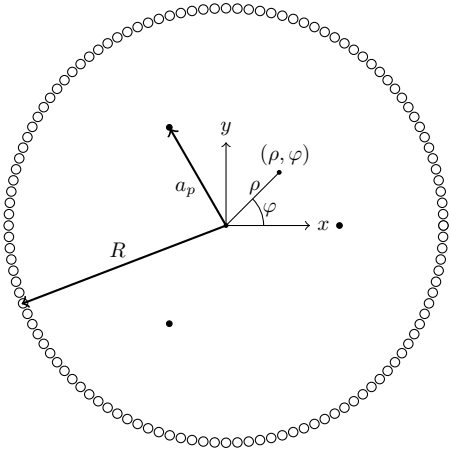
To investigate and focus on the armour losses, we study an un-screened three-core cable, with the core conductors represented as filamentary conductors. In \cite{Haber1974} expressions for the magnetic field surrounding three twisted filamentary conductors carrying balanced three-phase currents are derived. Let the global cylindrical coordinate system be such that the positive $z$-axis points along the cable's main axis, and that the $xy$-plane is placed in the cable cross section as depicted in Fig. \ref{fig:cable_fig_1}. The expression for the $z$-component of the auxiliary field found by \cite{Haber1974} is quite lengthy, and a more compact but equivalent representation is
\begin{equation} \label{eq:H0z}
    H_{e, z} (\rho, \varphi, z) = \sum_{m=-\infty}^\infty E_m K_m(\eta_m \rho) e^{jm(\varphi - \frac{2\pi z}{p_c})}
\end{equation}
where $K_m(\eta_m \rho)$ is the modified Bessel function of the second kind of order $m$, $\eta_m = 2\pi |m| / p_c$ and $p_c$ the pitch of the helix traced out by the three filamentary conductors. The coefficient $E_m$ is given as
\begin{equation}\label{Em}
E_m = I_{c} \frac{1}{p_c} ( \eta_m a_p I_{m+1}(\eta_m a_p)+mI_{m}(\eta_m a_p))f(m)
\end{equation}
where $I_c$ is the magnitude of the core current, $I_{m/m+1}(x)$ is the modified Bessel functions of the first kind of order $m/m+1$ and $f(m) = 1 + e^{j2\pi m} + e^{-j2\pi m}$, i.e. 0 or 3 depending on the value of $m$. Lastly, $a_p$ is the distance from the center of the cable to the center of the three filamentary conductors.

The reader may wonder why only the $z$-component of the auxiliary field is introduced, to which the answer is that the three field components are related to each other by simple algebraic or differential relations. These relations will be introduced in section IV out of necessity. 

\section{Transformation of the $N$-wire armour to an equivalent tube}
To be able to tackle the problem analytically, it is found necessary to represent the wire armour as an equivalent tube. Thus, the aim of this section is to establish the parameters of the equivalent tube. The tube is equivalent in the sense that when placed in a uniform external magnetic field with a given strength and direction, it yields the same complex loss as the $N$ wire armour placed in the same field.

The tube is assumed to be non-conducting as no net current will flow in each armour wire, only eddy currents, and thus the main parameter to be established for the tube is its complex permeability. The imaginary part will then account for both hysteresis and eddy current losses. As the wire armour's response is dependent on the direction of an applied uniform magnetic field, it will necessarily require the equivalent tube to be anisotropic.

The components of the effective permeability tensor $\mu'$ of the tube representation is derived in the (curvilinear) principal axes system of the tube, denoted by primed coordinate variables. In the principal coordinate system the $z'$ coordinate curve follows the helical path taken by the armour wires which are twisted with a pitch angle $\theta_a$, or equivalently, laid with a pitch $p_a$. The $\rho'$ basis vector coincides with $\rho$ in the global coordinate system, and the $\varphi'$ basis vector is everywhere perpendicular to both the $z'$ and $\rho'$ basis vectors. By definition of the principal axes system, the effective permeability is diagonal with $\mu' = \mathrm{diag}(\mu_{\rho'}, \mu_{\varphi'}, \mu_{z'})$. However, as the principal axes system can be understood to be a coordinate system that is rotated an angle $\theta_a$ around the $\rho$-axis from the global coordinate system introduced earlier, the effective permeability tensor expressed in the global system will not be diagonal. In fact, in the global coordinate system the permeability tensor is equal to $\mu = \mathbf{R} \mu' \mathbf{R}^t$, where $\mathbf{R}$ is a rotation matrix which governs the mentioned rotation around the $\rho'$-axis. Hence, \break \vspace{-6mm}

\begingroup
\footnotesize
\begin{equation}\label{mutensor}
\begin{split}
    \mu &= \begin{bmatrix}
        \mu_{\rho\rho} & 0 & 0 \\
        0 & \mu_{\varphi \varphi} & \mu_{\varphi z} \\
        0 & \mu_{\varphi z}  & \mu_{zz}
    \end{bmatrix} \\
    &= \begin{bmatrix}
        \mu_\rho & 0 & 0 \\
        0 & \mu_{\varphi '} \cos^2{\theta_a}  + \mu_{z '} \sin^2{\theta_a} & \cos{\theta_a} \sin{\theta_a} (\mu_{z'}-\mu_{\varphi '}) \\
        0 & \cos{\theta_a} \sin{\theta_a} (\mu_{z'}-\mu_{\varphi '}) & \mu_{z'} \cos^2{\theta_a} + \mu_{\varphi '} \sin^2{\theta_a}
    \end{bmatrix}.
\end{split}
\end{equation}
\endgroup
The task at hand is then to determine two of the components of $\mu'$, namely $\mu_{\varphi'}$ and $\mu_{z'}$. The last component $\mu_\rho$ is found to be superfluous for the model derived herein.

To determine these tensor components it is thus necessary to find the field or vector potential solutions for both the wire representation and the tube representation in the uniform field. From the solutions, the complex power per meter $\Delta S$ dissipated by either representation can be calculated. For the tube representation, the power $\Delta S$ will be a function of its permeability, which can thus be solved for by equating the losses, i.e. $\Delta S_{wires} = \Delta S_{tube} (\mu_i)$.

In general, the solutions for the wires are found by solving the Helmholtz equation for the magnetic vector potential under the Coulomb gauge and assuming the validity of the constitutive relation $\mathbf{B}=\mu \mathbf{H}$, i.e.
\begin{equation}\label{VPoisson}
    \nabla^2 \mathbf{A} = - \mu \mathbf{J}
\end{equation}
where the source term on the r.h.s. depends on which domain the equation is to be solved in. Outside the armour wires the current density is zero and for the equivalent tube the current density is zero everywhere as the tube is non-conducting. Inside the wires, Ohm's law $\mathbf{J}=\sigma \mathbf{E}$ and using the expression of the electric field $\mathbf{E} = -\nabla \phi - j\omega \mathbf{A}$ gives
\begin{equation}\label{VHelmholtz}
    \nabla^2 \mathbf{A} = \mu \sigma j\omega \mathbf{A} = \kappa^2 \mathbf{A}
\end{equation}
when setting the scalar potential $\phi = 0$ everywhere. This choice of $\phi$ poses no issues as we are dealing with domains in which the electric field points in directions of constant conductivity $\sigma$, which leaves some residual gauge freedom within the Coulomb gauge $\nabla \cdot \mathbf{A} = 0$. 

The above equations, together with the appropriate boundary and interface matching conditions, allows us to determine the magnetic vector potential solutions, losses and finally the tube transformation.

\subsection{Longitudinal external field in the $z'$-direction}
The problem of establishing an effective permeability for wires placed in a uniform external field $\mathbf{H}_e = H_e \mathbf{a}_{z'}$ is eased greatly upon realizing that each wire neither affects the field outside itself, and thus nor the field in other wires. The problem has been solved earlier by one of the authors, as well as other authors \cite{Hatlo2015, Goddard2015}. We simply restate the result herein, with the equivalent permeability in the $z'$-direction given as
\begin{equation}
    \mu_{z'} = \frac{2\mu I_1(\kappa r)}{\kappa r I_0(\kappa r)}.
\end{equation}
where $r$ is the wire radius, and $I_{0}(\kappa r)$ and $I_{1}(\kappa r)$ are the modified Bessel functions of the first kind of zeroth and first order, respectively.
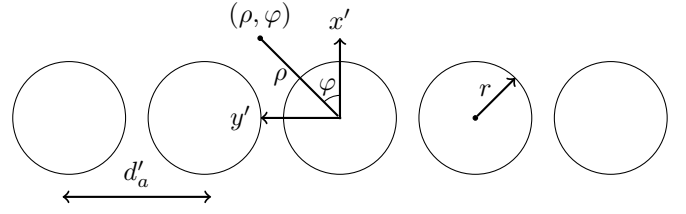
\begin{figure}[ht]
    \centering
    \begin{tikzpicture}[scale=0.3, every node/.style={scale=1}]
\node (O) at (0, 0) {};
\node (X) at (0, 4.4cm) {$x'$};
\node (Y) at (-4.4cm, 0) {$y'$};

\def \R {2.5cm};
\def \da {6cm};

\node (neg1O) at (\da, 0) {};
\node (neg1r) at (\da+1.7677cm, 1.7677cm) {};

\foreach \x in {-2, -1, 0, 1, 2}{
    \draw[thin] (\da*\x, 0) circle [radius=\R];
}

\draw[thick, ->, shorten <=-4pt] (O) -- (X);
\draw[thick, ->, shorten <=-4pt] (O) -- (Y);

\draw[thick, -, shorten <=-4pt] (O) -- node [scale=1, midway, below, yshift=0.2cm, xshift=-0.2cm] {$\rho$} (135:5cm);
\filldraw[] (135:5cm) circle [radius=0.1cm] node[above]{$(\rho, \varphi)$};
\draw (135:1cm) arc (135:90:1cm) node[midway, above, xshift=-0.05cm, yshift=-0.1cm]{$\varphi$};

\draw[thick, <->, shorten <=-2.5pt, shorten >=-2.5pt] (-2*\da, -\R-1cm) -- node [scale=1, midway,above] {$d_a'$} (-1*\da, -\R-1cm);

\filldraw[] (neg1O) circle [radius=0.1cm];
\draw[thick, ->, shorten <=-4pt, shorten >= -5pt] (neg1O) -- node [scale=1, midway, above, yshift=-0.1cm, xshift=-0.13cm] {$r$} (neg1r);

\end{tikzpicture}
    \caption{Local coordinate system for a string of infinitely many wires with radius $r$ spaced a distance $d_a'$ apart.}
    \label{fig:wirestring}
\end{figure}
\vspace{-5mm}
\subsection{Transverse external field in the $\varphi '$-direction}
For a transverse external field, the armour wires will alter the field outside and are not decoupled. Instead of studying a ring of $N$ armour wires we examine an infinite string of armour wires $k=-\infty, ..., -1, 0, 1, ..., \infty$, similar to the approach in \cite{Goddard2015,Hatlo2015}. We introduce a local coordinate system with wire $k=0$ located at the origin, and with coordinate axes and coordinate variables as shown in Fig. \ref{fig:wirestring}. Let the string of wires be placed in a uniform external field $\mathbf{H}_e$ in the -$y'$-direction. A choice for the corresponding magnetic vector potential is
\begin{equation}
    \mathbf{A}_e(\rho, \varphi) = \mu_0 H_e \rho \cos{\varphi} \  \mathbf{a}_{z'}.
\end{equation}
The reponse of the wires will only cause a magnetic vector potential in the $z'$-direction. Thus, \eqref{VPoisson} reduces to the scalar Laplace equation outside the wires, and \eqref{VHelmholtz} reduces to the scalar Helmholtz equation inside the wires. Linearity of Maxwell's equations, combined with symmetry and finiteness considerations, causes the general solution outside the wires, denoted by a superscript (o), to be on the form
\begin{equation} \label{GeneralAzo}
    A_{z'}^{(o)}(\rho, \varphi) = \mu_0 H_e \rho \cos{\varphi} + \sum_{k=-\infty}^{\infty} \sum_{m=1}^\infty D_m \rho_k^{-m} \cos{m \varphi_k}
\end{equation}
where $\rho_k$ and $\varphi_k$ are referred to the center axis of the $k$-th wire, and $D_m$ are the undetermined coefficients of the general solution. Each wire is thus exposed not only to the external field, but also to the response from all other wires. The issue of determining the coefficients $D_m$ is not resolved without referring the variables $\rho_k$ and $\varphi_k$ to a common wire axis. Taking the same approach as Carson \cite{Carson1921, Carson1922}, the quantity  $\rho_k^{-m} \cos{m\varphi_k}$ can be referred to the axis of wire $k=0$ and expressed solely by the variables $\rho$ and $\varphi$ through the transformation
\begin{equation}
    \frac{\cos{m\varphi_k}}{\rho_k^m} = \sum_{n=0}^{\infty}\frac{(-1)^n}{(k d_a)^m} \frac{m^{\Bar{n}}}{n!} \big(\frac{\rho}{k d_a}\big)^n \cos(n \varphi)
\end{equation}
where $m^{\Bar{n}}$ denotes the rising factorial $m(m+1)(m+2)...(m+n-1)$ and $d_a'$ is the wire spacing as shown in Fig. \ref{fig:wirestring}. The transformation is valid and converges close to wire zero, i.e. for $\rho / (k d_a) < 1$. Thus, by substitution the solution is on the form
\begin{equation} \label{eq:Azo}
\begin{split}
A_{z'}^{(o)}(\rho, \varphi) = \mu_0 H_e \rho \cos{\varphi} + \sum_{m=1}^\infty D_m \big[\rho^{-m} \cos{m \varphi} \ + \\ \sum_{n=0}^{\infty}\frac{(-1)^n}{d_a^{m+n}} \frac{m^{\Bar{n}}}{n!} \zeta(m+n) (1 + (-1)^{m+n}) \rho^n \cos(n \varphi) \big]
\end{split}
\end{equation}
which is arrived at by interchanging summations, re-writing the sum over wires so that $k=1...\infty$ and recognizing $\sum_{k=1}^\infty k^{-m-n}$ as the Riemann-Zeta function $\zeta(m+n)$. Inside wire $k=0$, the general solution, denoted by superscript $(i)$, is
\begin{equation} \label{eq:Azi}
    A_{z'}^{(i)}(\rho, \varphi) = \sum_{m=1}^\infty A_m I_m(\kappa \rho) \cos{m \varphi}.
\end{equation}
where $A_m$ are the undetermined coefficients and $I_m(\kappa \rho)$ is the modified Bessel function of the first kind of order $m$. The coefficients $A_m$ and $D_m$ can now be determined by matching the internal and external solutions at the wire surface by enforcing continuity of $B_\rho$ and $H_\varphi$. Applying these interface matching conditions leads to the equations
\begin{align}
    D_m &= A_m q_m r^m \label{TransverseCoefficientDm}, \\
    A_m &= \frac{1}{p_m} \big[\mu_0 H_e r \delta_{m1}+ r^m \sum_{n=1}^\infty  A_n q_n r^n  f_{mn} \big], \label{TransverseCoefficientAm}
\end{align}
for the undetermined coefficients, where $\delta_{m1}$ is the Kronecker delta, $r$ the wire radius and
\begin{align}
    q_m &= \frac{1}{2}\big( I_m(\kappa r) - \frac{\mu_0 \kappa r}{2\mu m}(I_{m-1}(\kappa r) + I_{m+1}(\kappa r))\big), \label{eq:qm} \\
    p_m &= \frac{1}{2}(I_m(\kappa r) + \frac{\mu_0 \kappa r}{2\mu m}(I_{m-1}(\kappa r) + I_{m+1}(\kappa r)), \\
    f_{mn} &= \frac{(-1)^n}{d_a^{m+n}}\frac{m^{\Bar{n}}}{n!}\zeta(m+n)(1+(-1)^{m+n}).
\end{align}
$q_n$ appearing in the sum of \eqref{TransverseCoefficientAm} is also given by \eqref{eq:qm}, by letting $m \to n$. Equations \eqref{TransverseCoefficientDm} and \eqref{TransverseCoefficientAm} form a system of an infinite number of linear equation for the infinite number of unknowns $A_m$ and $D_m$. Truncating this system to one unknown allows for approximating its solution as
\begin{align}
    A_1 &= \frac{\mu_0 H_e r}{p_1 - r^2 q_1 f_{11}}, \label{eq:A1}\\
    D_1 &= \mu_0 H_e \frac{q_1 r^2}{p_1 - r^2 q_1 f_{11}}, \label{eq:D1}
\end{align}
which is accurate as long as $f_{1n}r^{n+1}\propto (r/d_a)^{n+1}<<1$ for $n> 1$.

The specific solution for the vector potential has now been obtained, which can be used to compute the wire losses. However, some "trick" is needed to compute the losses caused by $N$ wires from the solution above, which is valid for an infinite string of wires. The "trick" and the way forward is to  \textit{mathematically} decouple the wires by seeing that the contribution to the total field from a single wire can simply be found by keeping the term $k=0$ in \eqref{GeneralAzo}
\begin{equation}
    \mathbf{A}^{(o)}(\rho, \varphi) = (\mu_0 H_e \rho + D_1 \frac{1}{\rho}) \cos{\varphi} \ \mathbf{a}_{z'}.
\end{equation}
This vector potential is the sum of contributions from the external field and the response field due to a single wire, but the wire's response takes into consideration that it is placed in a string of infinitely many other wires through $D_1$. To avoid evaluating volume integrals over all space to find the complex losses caused by the wires, a specific power loss integral based on the application of Poynting's theorem is derived in Appendix A. The apparent power due to $N$ armour wires is computed by this power loss integral, and is simply equal to $j\omega 2 \pi N D_1 H_e$.

Just as the field solution and complex losses was established for the string of armour wires, the same is necessary to do for the non-conducting equivalent tube to establish its effective permeability. The tube is modelled as an infinite strip along the $y'$-direction, with a thickness $t$. The magnetic field is constant within the infinite strip for a uniform external field in the $-y'$-direction, with the magnetic field $\mathbf{B}^{(i)}$ being equal to $\mu_{\varphi'} \mathbf{H}_e$ and $\mathbf{H}^{(i)} = \mathbf{H}_e$. The power dissipated by a finite section of the tube can be neatly calculated as an integral over the volume of section $v_t$ as
\begin{align} \label{eq:jackson}
    \Delta S &=j\omega \int_{v_{t}} \mathbf{B}^{(i)}\cdot\mathbf{H}_e^* - \big(\mathbf{H}^{(i)}\big)^* \cdot \mathbf{B}_e dv \\ &= j \omega (\mu_{y'} - \mu_0) |\mathbf{H}_e|^2 \int_{v_{t}} dv
\end{align}
with \eqref{eq:jackson} derived in \cite{JacksonElectrodynamics}. Choosing a finite section of the infinite strip with the same volume as $N$ wires and equating the dissipated powers, we arrive at the expression for the effective permeability $\mu_{\varphi'}$,
\begin{equation} \label{eq:muvarphi}
    \mu_{\varphi'} = \mu_0 \big[ 1 + \frac{2 q_1}{p_1 - q_1 r^2 f_{11}}\big].
\end{equation}
Before proceeding to the next section, one remark should be made about the parameter $d_a'$ occuring in \eqref{eq:muvarphi}. $d_a'$ in the local coordinate system is related to the wire spacing in the global coordinate system $d_a$ by
\begin{equation}
    d_a' = \frac{d_a}{\sqrt{1+\big(\frac{2\pi R}{p_a}\big)^2}} = \frac{1}{N}\frac{2\pi R}{\sqrt{1+\big(\frac{2\pi R}{p_a}\big)^2}}.
\end{equation}
where $R$ is the mean radius of the armour layer and $p_a$ is the pitch of the armour wires.

\section{Armoured three-core cable}
Having established the effective permeability tensor $\mu$ in the previous section, our goal is now to find an expression for the losses occurring in the equivalent tube. The tube will no longer be placed in a uniform external field, but rather in the exciting field generated by three twisted filamentary conductors carrying balanced three-phase currents, as introduced in section II.

Let the equivalent tube armour of mean radius $R$ and thickness $t$ enclose the filamentary conductors. We operate in the global coordinate system, and divide the domain into three natural sub-regions,
\begin{align*}
    \mathrm{region \ 1} &= \{\rho \ |\ \rho < R - t/2\},  \\
    \mathrm{region \ 2} &= \{\rho \ |\ R - t/2 < \rho < R + t/2\},  \\
    \mathrm{region \ 3} &= \{\rho \ |\ \rho > R + t/2\},
\end{align*}
with the fields in each region denoted by superscript corresponding to the region number in the upcoming text. Under the tube transformation described in the previous section, the equivalent tube carries no free currents. Then, the magnetic field $\mathbf{B}^{(2)}$ and the accompanying auxiliary field $\mathbf{H}^{(2)}$ in the tube are governed by the two source-free Maxwell's equations
\begin{IEEEeqnarray}{rCL} \label{eq:CurlH}
\nabla \times \mathbf{H}^{(2)} = 0, \\ \label{eq:DivB}
\nabla \cdot \mathbf{B}^{(2)} = 0.
\end{IEEEeqnarray}
The three components of Maxwell's curl equation \eqref{eq:CurlH} are
\begin{IEEEeqnarray}{rCL} \label{eq:CurlH1}
\frac{jm}{\rho} H_{z}^{(2)} + jm\frac{2\pi}{p_c} H_{\varphi}^{(2)} = 0 \\
\frac{jm2\pi}{p_c} H_\rho^{(2)} + \frac{\partial H_{z}^{(2)}}{\partial \rho} = 0 \\ \label{eq:CurlH3}
\frac{\partial (\rho H_{\varphi}^{(2)})}{\partial \rho} - jm H_\rho^{(2)} = 0
\end{IEEEeqnarray}
for the $m$-th term of an infinite series solution. The derivatives w.r.t. $z$ and $\varphi$ have here been reduced to algebraic factors by utilizing periodicity of the auxiliary field w.r.t. these variables. In the same manner, the $m$-th term of the divergence equation \eqref{eq:DivB} can be expressed as
\begin{equation}
\begin{split}
    \frac{1}{\rho}\frac{\partial(\rho \mu_{\rho} H_\rho^{(2)})}{\partial \rho} + \frac{j m}{ \rho}(\mu_{\varphi} H_{\varphi}^{(2)} + \mu_{z\varphi} H_z^{(2)}) \\- \frac{jm2\pi }{p_c} (\mu_{z\varphi} H_\varphi^{(2)} + \mu_{z} H_z^{(2)}) = 0
\end{split}
\end{equation}
where $\mu_{z\varphi}$ are the off-diagonal terms in the permeability tensor $\mu$ as given by \eqref{mutensor}. Substituting in the appropriate relations between components from \eqref{eq:CurlH1} - \eqref{eq:CurlH3} yields an ODE for the $H_z^{(2)}$ component, which after some manipulations becomes
\begin{equation}
\begin{split}
\label{eq:odewhittaker}
    \rho\frac{\partial}{\partial \rho}\big(\rho \frac{\partial (\mu_{\rho \rho} H_z^{(2)})}{\partial \rho}\big) - \\ \Big( \big(\frac{2\pi m}{p_c}\big)^2 \mu_{zz} \rho^2 - m^2 \frac{4\pi}{p_c} \mu_{z\varphi} \rho + m^2 \mu_{\varphi \varphi}\Big) H_z^{(2)} = 0.
\end{split}
\end{equation}
While the exact solution to the above ODE is given by Whittaker functions, we instead look for a simpler approximate solution by treating the equivalent tube as a thin shell with specific matching conditions between the inner and outer surface of the shell, directly connecting region 1 and 3, and thus eliminating region 2. To achieve this we make use of the fact that the tube thickness $t$ is small compared to the tube radius $R$, allowing for assuming $\rho \approx R$ across the tube. In addition, we assume that the field $H_z^{(2)}$ is constant across the tube. The above ODE simplifies to
\begin{equation} \label{eq:ODEHzsimp}
    \frac{\partial}{\partial \rho}\frac{\partial (\mu_{\rho \rho} H_z^{(2)})}{\partial \rho}- \mu_e \eta_m^2 H_z^{(2)} = 0
\end{equation}
where $\mu_e = (\mu_{\varphi \varphi} \big(\frac{p_c}{2 \pi R}\big)^2 - \frac{2p_c}{2 \pi R}\mu_{z\varphi} + \mu_{zz})$. Multiplying both sides by $\partial \rho$ and integrating from the inner to the outer surface of the armour yields the first matching condition at $\rho = R$,
\begin{equation} \label{eq:matchingcondition1}
    \frac{\partial H_z^{(3)}}{\partial \rho}-\frac{\partial H_z^{(1)}}{\partial \rho} = \mu_e \eta_m^2 H_z^{(2)}(R) t.
\end{equation}
When going from \eqref{eq:ODEHzsimp} to \eqref{eq:matchingcondition1} $\mu_{\rho \rho}$ simply becomes 1 as the integral limits are assumed to be in region 1 and 3. 

The aforementioned assumption that the field $H_z^{(2)}$ does not change significantly across the armour tube is exactly the second matching condition
\begin{equation} \label{eq:matchingcondition2}
    H_z^{(3)} - H_z^{(1)} = 0.
\end{equation}
For the magnetic field in region 1 and 3, we introduce the exciting field $\mathbf{H}_e$ generated by the current flowing in the filamentary conductors as defined in Section II. A response field $\mathbf{H}_r$ occurs due to the presence of the equivalent tube armour, and the total field can be represented by a linear combination of the two. The source-free Maxwell's equations for the total field $\mathbf{H} = \mathbf{H}_e + \mathbf{H}_r$ reduces to
\begin{IEEEeqnarray}{rCL}
\nabla \times \mathbf{H} = \nabla \times \mathbf{H}_r = 0, \\
\nabla \cdot \mathbf{B} = \nabla \cdot \mathbf{B}_r = 0
\end{IEEEeqnarray}
as the exciting field \eqref{eq:H0z} already solves these equations. 

The ODE for the $z$-component of the auxiliary field $\mathbf{H}_r$ is found by letting $\mu_{\rho \rho} = \mu_{\varphi \varphi} = \mu_{z z} = \mu_0$ and $\mu_{z\varphi} = 0$ in \eqref{eq:odewhittaker}, reducing it to the modified Bessel's equation. Due to finiteness considerations, the $m$-th term in the general solutions for the $z$-component becomes
\begin{align}
    H_{z, m}^{(1)} = B_m I_m(\eta_m \rho) + H_{e,  z, m}, \\
    H_{z, m}^{(3)} = C_m K_m(\eta_m \rho) + H_{e,  z, m}.
\end{align}
The specific solution can now be found, as the undetermined coefficients $B_m$ and $C_m$ in the above solutions in region 1 and 3 are determined by the matching conditions \eqref{eq:matchingcondition1} and \eqref{eq:matchingcondition2}. Enforcing these causes the $m$-th term of the specific solution for the auxiliary field inside the tube to be
\begin{equation}\label{eq:tubesol}
    H_{z, m}^{(2)}(R) = \frac{H_{e, m}}{1+\frac{\eta_m^2 \mu_e t}{\xi_m}}
\end{equation}
where $\xi_m = \frac{I_m'}{I_m}-\frac{K_m'}{K_m}   = \frac{1}{R I_m(\eta_m R) K_m(\eta_m R)}$.
Since the denominator is positive, the auxiliary field is reduced compared to an unarmoured cable.  

The apparent power dissipated by the equivalent tube due to the enclosed filamentary conductors can be found by the integral
\begin{equation}
    \Delta S = j\omega \int_{v_t} \mathbf{B}\cdot \mathbf{H}_e^* - \mathbf{H}^*\cdot \mathbf{B}_e dv
\end{equation}
over the volume $v_t$ of the tube \cite{JacksonElectrodynamics}. Taking into account the an-isotropic permeability of the tube, the above solution for the (constant) field inside the tube and utilizing the relations between the field components, the apparent equivalent tube loss is
\begin{equation}\label{eq:loss}
    \Delta S = j\omega 2 \pi R t \mu_0 \sum_{m=-\infty}^\infty \frac{\mu_e - \big(\frac{p}{2 \pi R}\big)^2 - 1}{1+\frac{\eta_m^2 \mu_e t}{\xi}} |H_{m, e, z}|^2 .
\end{equation}
Lastly, the reader should note that the tube thickness $t$ occurring in \eqref{eq:tubesol} and \eqref{eq:loss} is not equal to the wire diameter, but equal to
\begin{equation}
    t = \frac{N r^2}{2 R}  \sqrt{1 + \Big(\frac{2\pi R}{p_a}\Big)^2}
\end{equation}
such that the volume of the equivalent tube is equal to the volume of the armour wires.

Using this derived expression of the armour losses the armour loss factor as defined in IEC 60287 can be calculated with the formula  

\begin{equation}
    \lambda_2 = \Re\Bigg{(}\frac{j\omega 2 \pi R t \mu_0}{3R_{AC}} \sum_{m=-\infty}^\infty \frac{\mu_e }{1+\frac{\eta_m^2 \mu_e t}{\xi}} |H_{m, e, z}|^2 \Bigg{)}.
\end{equation}
Valid for magnetic armour where
\begin{equation}
    \mu_e\gg\big(\frac{p}{2 \pi R}\big)^2 +1:
\end{equation}

This formula may be directly used in the IEC 60287 formalism.  
\begin{table}[!t]
\renewcommand{\arraystretch}{1.3}
\caption{Example cable design data.}
\label{table:cabledata}
\centering
\begin{tabular}{c||c||c}
\hline
\bfseries Parameter & \bfseries Description & \bfseries Value\\
\hline\hline
$a_p$ & Core helix radius & 52.25 mm \\
$R$ & Mean radius of armour layer & 115.6 mm \\
$r$ & Armour wire radius & 2.5 mm \\
$\sigma$ & Armour wire conductivity & 5.3763 MS/m \\
$\mu_r$ & Armour wire rel. permeability & 150-j50, 600-j350 \\
$p_a$ & Pitch of armour wires & -100 m, -4 m, -2.4 m \\
$p_c$ & Pitch of cores & 1.2 m, 4 m, 2.4 m\\
$N$ & No. of armour wires & $\in [25, 135]$ \\
$w$ & Electrical angular frequency & 314.16 rad/s \\
$I_c$ & Core current & 1000 A \\
\hline
\end{tabular}
\end{table}
\begin{figure}[ht]
    \centering
    \includegraphics[trim={0 0 0 0}, width=0.45\textwidth]{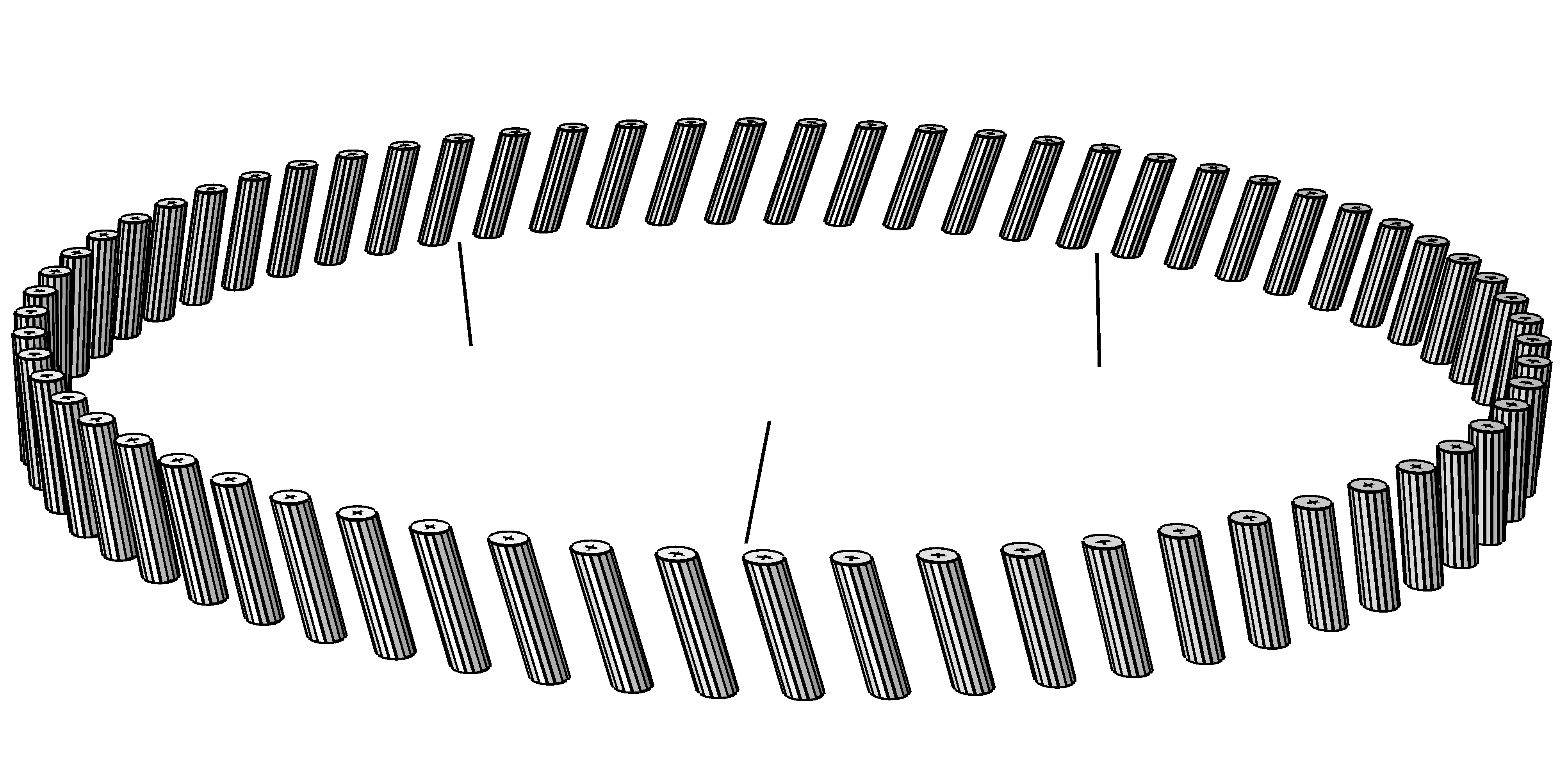}
    \caption{The modified COMSOL model based on \cite{comsoltutorial}, with three twisted edge current enclosed in a ring of twisted armour wires.}
    \label{fig:comsolmodel}
\end{figure}
\section{Model validation}
The proposed armour loss formulation (\ref{eq:loss}) is compared with results from FEA in COMSOL Multiphysics 5.6. The COMSOL model used herein is a modified version of the COMSOL Cable Tutorial Series' short twisted-periodicity model \cite{comsoltutorial}. The modification mainly consists of replacing the core conductors and sheaths with twisted filamentary conductors carrying edge currents, in addition to mesh refinement in and in vicinity of the armour wires. The mesh refinement was found necessary to achieve a satisfactory accurate result. The COMSOL model is depicted in Fig. \ref{fig:comsolmodel}.
\begin{figure}[ht]
    \centering
    \begin{subfigure}[t]{0.46\textwidth}
        \includegraphics[trim={0 0.7cm 0 0}, width=\textwidth]{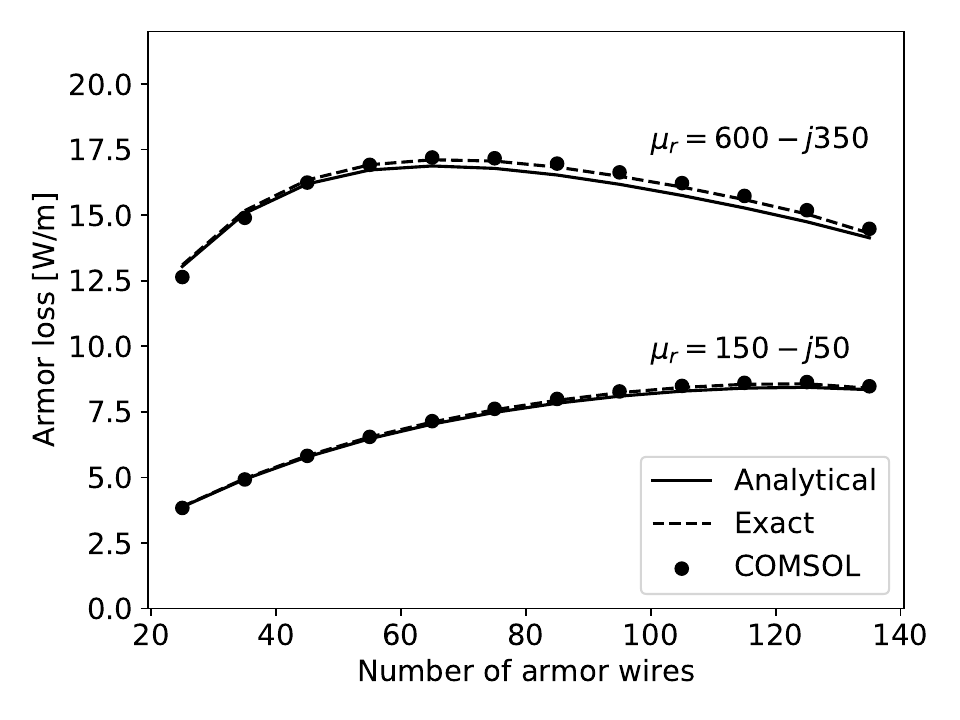}
        \caption{$p_c = 1.2$ m, $p_a = -100$ m.}
        \label{fig:result_fig_1_a}
    \end{subfigure}
    \vskip 0cm 
    \begin{subfigure}[t]{0.46\textwidth}
        \includegraphics[trim={0 0.7cm 0 0}, width=\textwidth]{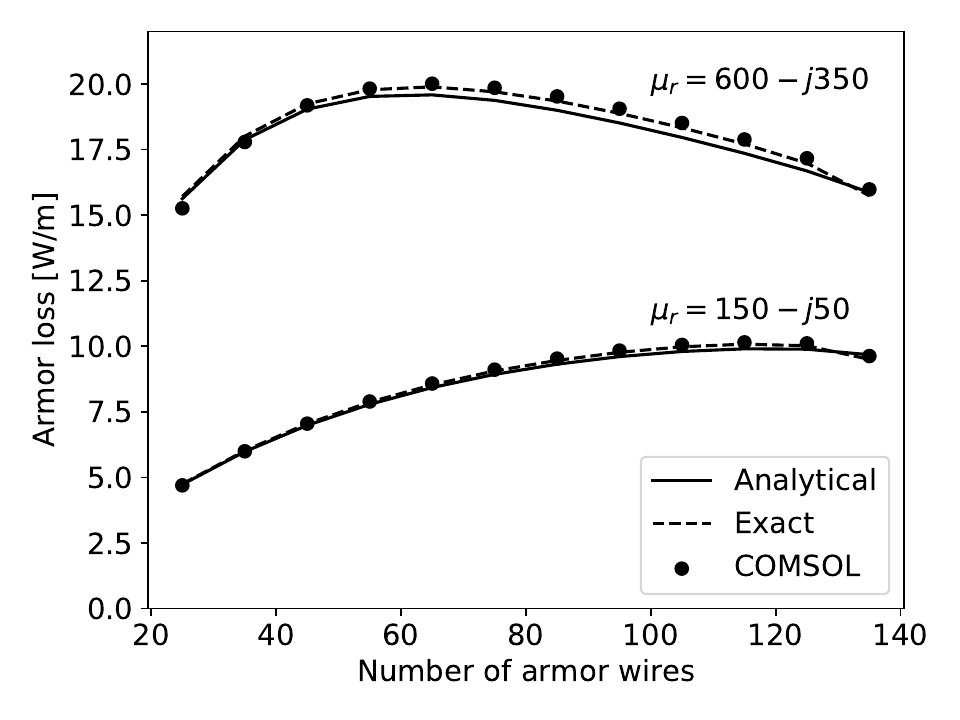}
        \caption{$p_c = 2.4$ m, $p_a = - 2.4$ m.}
        \label{fig:result_fig_1_b}
    \end{subfigure}
    \vskip 0cm 
    \begin{subfigure}[t]{0.46\textwidth}
        \includegraphics[trim={0 0.7cm 0 0}, width=\textwidth]{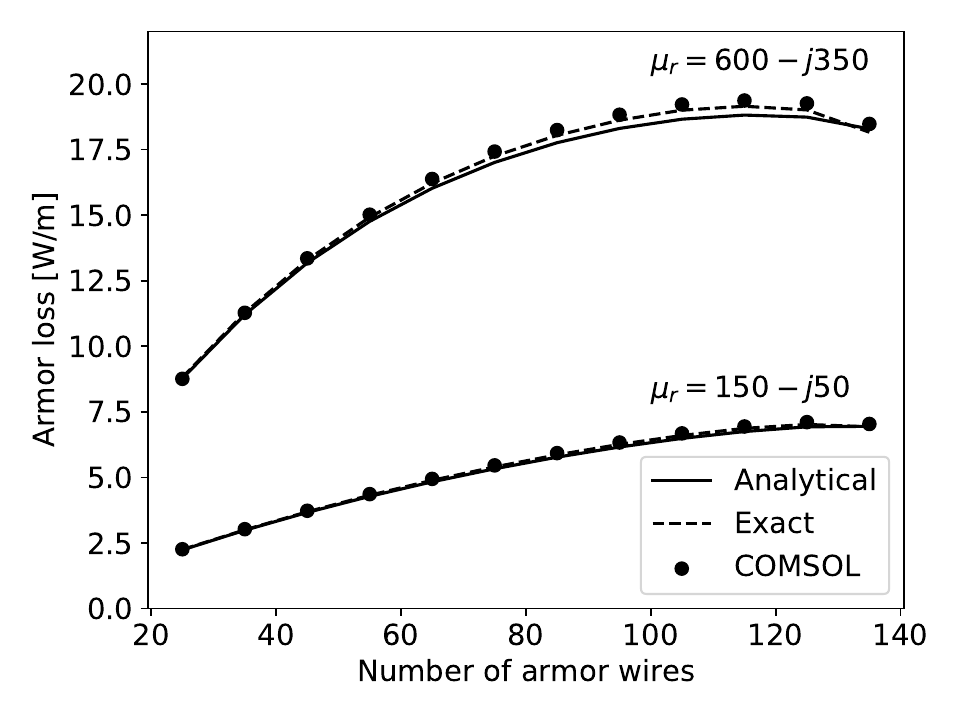}
        \caption{$p_c = 4$ m, $p_a = - 4$ m.}
        \label{fig:result_fig_1_c}
    \end{subfigure}
    \caption{Armour losses for different combinations of core and armour wire pitches and relative permeabilities, computed by (\ref{eq:loss}) as functions of number of armour wires.}
    \label{fig:result_fig_1}
\end{figure}

The example cable used for validation purposes have the design data listed in Table \ref{table:cabledata}. The pitch length of the cores and armour wires, as well as the armour wire steel permeability and the number of armour wires are varied.

The solid line in Fig. \ref{fig:result_fig_1} shows the armour losses computed as a function of the number of armour wires $N$ by the proposed (\ref{eq:loss}), and compared to the FEA results represented by dots. Different pairs of armour wire pitch and core pitch are used, and for each pairing two values for the armour wire complex permeability $\mu_r$ are used.

While the approximate analytical solution is the main result of this paper, an exact solution was derived for comparison purposes. The exact solution differs from the approximate solution by (i) utilizing Whittaker functions for the field solution within the tube, and (ii) truncating the linear system defined by \eqref{TransverseCoefficientAm} to 17 unknowns and 17 equations, instead of one unknown and one equation as per \eqref{eq:A1} and \eqref{eq:D1}. The exact solution is shown as the dashed lines in Fig. \ref{fig:result_fig_1}.

Both the approximate analytical solution and the exact solution are in good agreement with the COMSOL results. For the cases studied, a maximum deviation of 3.2 \% between the approximate analytical solution and COMSOL is observed, occurring for 25 armour wires. For a fully-armoured cable with 135 wires the maximum deviation is 2.4 \% for the analytical solution. For the exact solution, the maximum deviation for fully armoured cables is approximately 1.7 \%.
\begin{figure}[ht]
    \centering
    \begin{subfigure}[t]{0.46\textwidth}
        \includegraphics[trim={0 0.7cm 0 0}, width=\textwidth]{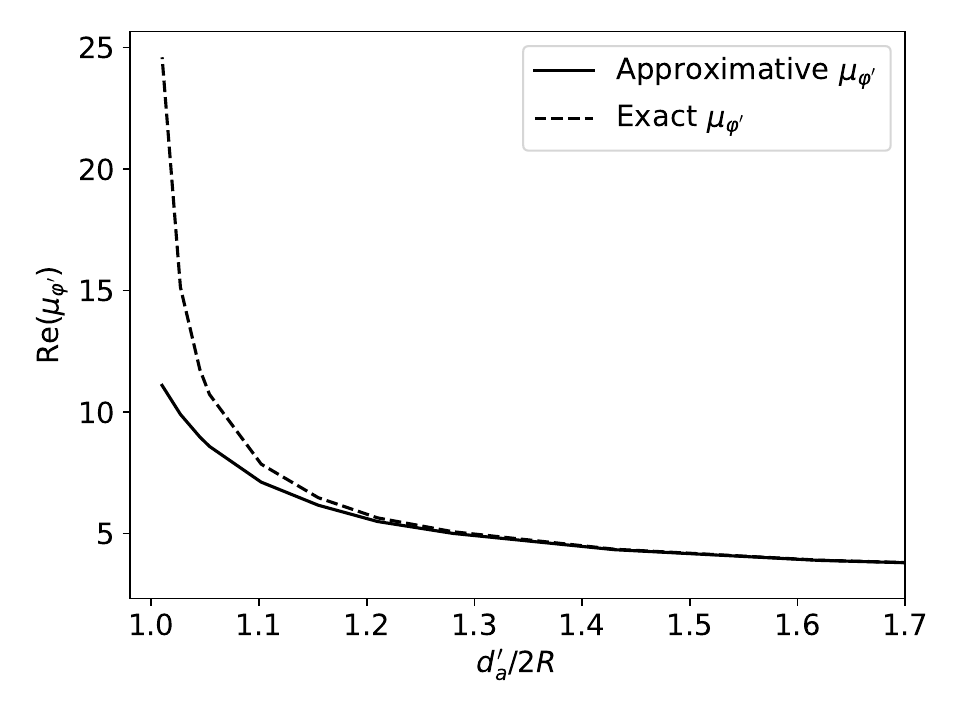}
        \caption{Real part of $\mu_{\varphi'}$.}
        \label{fig:result_fig_2_a}
    \end{subfigure}
    \vskip 0cm 
    \begin{subfigure}[t]{0.46\textwidth}
        \includegraphics[trim={0 0.7cm 0 0}, width=\textwidth]{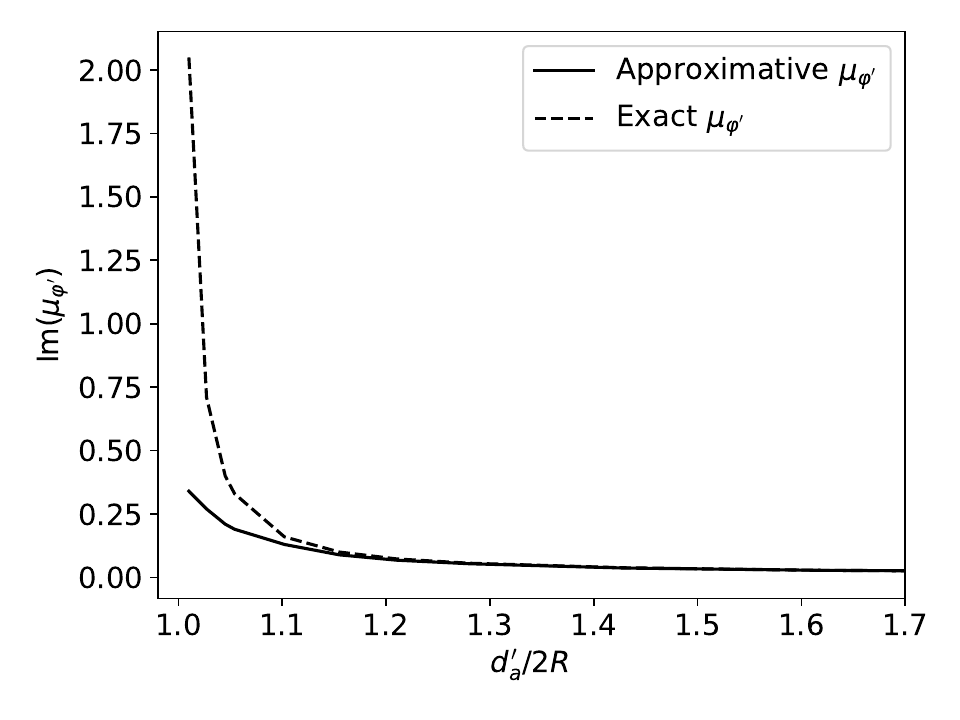}
        \caption{Imaginary part of $\mu_{\varphi'}$.}
        \label{fig:result_fig_2_b}
    \end{subfigure}
    \caption{Effective relative permeability of the armour layer in the $\varphi'$-direction as a function of relative wire spacing $d_a'/2R$. The solid lines are generated from \eqref{eq:muvarphi}, whilst the dashed lines are obtained after truncating the linear system of \eqref{TransverseCoefficientAm} to 17 unknowns and 17 equations.} \label{fig:result_fig_2}
    \vskip -0.4cm
\end{figure}
\section{Discussion}
One heuristic in the derivation of the proposed model is the transformation of the wire armour to an equivalent tube, which is made such that both armour representations yield the same complex loss when placed in a uniform external magnetic field of a given strength. After this initial tuning, the tube representation is used when the external field is no longer uniform, but instead generated by balanced three-phase currents carried by the filamentary conductors. All investigations performed suggest that this heuristic does not introduce significant errors for realistic cable designs. However, the largest deviations with COMSOL is observed for a low number of armour wires, likely due to the tube representation being quite unphysical for very few armour wires.

In addition to the above-mentioned heuristic, some central simplifying assumptions were made in the derivation of the model. Two assumptions helped simplify (\ref{eq:odewhittaker}) - it was assumed firstly that the armour tube thickness is insignificant compared to the armour radius (t/d <<1), and secondly, that the field inside the armour is approximately constant in the $\rho$-direction. A third simplifying approximation was the truncation to one unknown and one equation in the infinite system defined by (\ref{TransverseCoefficientAm}). These assumptions are good in the sense that the approximate solution only introduces a small error in comparison with the exact solution and the COMSOL computations, as can be seen in Fig. \ref{fig:result_fig_1}.

The number of unknowns included in the truncation of \eqref{TransverseCoefficientAm} affects the losses indirectly by affecting $\mu_{\varphi'}$ directly. A comparison of retaining one and 17 terms is shown in Fig. \ref{fig:result_fig_2}. A small difference between the two differently truncated systems are seen for relatively large wire-spacing $d_{a}'/2R$, but for high packing factors both the real and imaginary part of the permeability increases rapidly. While this approximation introduce large errors for the equivalent permeability $\mu_{\varphi'}$ for small gaps between adjacent armour wires, the effect on power dissipation is seemingly very small.

Although not a part of the main results, the authors have observed that mesh refinement studies are of crucial importance for these kinds of models - and meshes that were believed to resolve the underlying physics well enough, did not do so since a high degree of accuracy was necessary for model validation purposes. For a small selection of cases, meshes were refined even further than the meshes used to produce the main results herein. The mesh refinement on these few, selected cases seemed to have a non-negligible effect on computed losses, but at the cost of an explosive rise in computation time. The effect of refining the meshes further for these cases lead to lower losses computed by COMSOL.

\section{Conclusion}
The model proposed for calculation of armour losses in this paper is easy to use, with the final result being an analytical formula. The formula can be implemented in a spreadsheet calculation tool, or a programming language such as e.g. Python or MATLAB, and then be used by non-specialists. The infinite series in (\ref{eq:loss}) converges rapidly and only a few terms is necessary to achieve satisfactory accuracy. If needed, the formula can be cast into the framework of IEC 60287, expressed as the so-called $\lambda_2$ factor which describes the ratio of the armour losses to the core conductor losses. 

With an analytical formula it is easier to understand the most important design and material properties, which enables better design of submarine three-core cables. 

\appendices
\section{}
To find a useful expression for the power dissipated by the wires placed in a uniform field, we start out with the time-averaged Poynting's theorem for quasi-magnetostatic fields
\begin{equation}
\oint_S \mathbf{E}\times \mathbf{H}^* \cdot d\mathbf{a} = - \int_V \mathbf{E} \cdot \mathbf{J}^* dv - j\omega \int_V \mathbf{B} \cdot \mathbf{H}^* dv.
\end{equation}
Letting the surface $S$ enclosing the volume $V$ with the $N
$ wires go to infinity causes the surface integral to vanish as no radiation occurs for low frequencies. A local source capable of generating a uniform field is two infinite parallel current sheets, with injected uniform current density $\mathbf{J}_e$. Extracting this injected current density from the first integral on the r.h.s. of the above equation yields
\begin{equation}
    \int_V \mathbf{E}\cdot\mathbf{J}_e^* dv = - \int_V \mathbf{E}\cdot \mathbf{J}^* dv - j\omega \int_V \mathbf{B} \cdot \mathbf{H}^* dv.
\end{equation}
If the injected current density is kept equal both with and without the wires placed between the sheets, then the complex power dissipated by the armour wires is the difference
\begin{equation}
    \Delta S = j\omega \int_V (\mathbf{A}-\mathbf{A}_e)\cdot \mathbf{J}_e^* dv
\end{equation}
where $\mathbf{A}_e$ is the vector potential for the external field, and where the relation $\mathbf{E} = -j\omega \mathbf{A}$ have been used. As we are dealing with two infinite parallel current sheets carrying currents of equal magnitude in opposite directions, the source current density can be expressed as $\mathbf{J}_e = K_e(\delta(x-x_0) - \delta(x+x_0))\mathbf{a}_z$, where $K_e$ is the sheet current density in A/m, and $\delta(x \pm x_0)$ are Dirac delta functions located at $\mp x_0$. The field strength $H_e$ between the sheets is simply equal to the current density $K_e$ carried by each sheet, and hence by symmetry considerations
\begin{align}
    \Delta S &= j\omega 2 H_e^* \int_{-\infty}^\infty  (\mathbf{A}(x_0, y)-\mathbf{A}_e(x_0, y))\cdot \mathbf{a}_z dy \\
    &= j\omega 2 H_e^* \int_{-\infty}^\infty A_{r, z}(x_0, y)  dy
\end{align}
where the last line follows as the total vector potential $\mathbf{A}$ with wires placed between the infinite sheets can be expressed as a sum of the response potential $\mathbf{A}_r$ and the external potential $\mathbf{A}_e$.

\ifCLASSOPTIONcaptionsoff
  \newpage
\fi

\bibliographystyle{IEEEtran}

\bibliography{IEEEabrv, references}
\end{document}